\RequirePackage{fix-cm}
\documentclass[smallextended]{svjour3}       
\smartqed  
\usepackage{color}
\usepackage{graphicx}
\usepackage{amsmath}
\usepackage{booktabs}
\usepackage{url}

\begin{document}

\title{Identifying macroscopic features in foreign visitor travel pathways
}


\author{Tatsuro Kawamoto         \and
	Ryutaro Hashimoto
}


\institute{T. Kawamoto \at
              Artificial Intelligence Research Center, \\
  National Institute of Advanced Industrial Science and Technology, \\
  2-3-26 Aomi, Koto-ku, Tokyo, Japan \\
              \email{kawamoto.tatsuro@aist.go.jp}           
           \and
           R. Hashimoto \at
           School of Political Science and Economics Department of Political Science,  \\
           Waseda University, \\
  1-6-1 Nishi-Waseda, Shinjuku-ku, Tokyo, Japan \\
              \email{h.ryuutarou1234@toki.waseda.jp}           
}

\date{Received: date / Accepted: date}

\maketitle

\begin{abstract}
Human travel patterns are commonly studied as networks in which the points of departure and destination are encoded as nodes and the travel frequency between two points is recorded as a weighted edge. 
However, because travelers often visit multiple destinations, which constitute pathways, an analysis incorporating pathway statistics is expected to be more informative over an approach based solely on pairwise frequencies. 
Hence, in this study, we apply a higher-order network representation framework to identify characteristic travel patterns from foreign visitor pathways in Japan. 
We expect that the results herein are mainly useful for marketing research in the tourism industry.
\end{abstract}

\section{Introduction}\label{intro}
The number of foreign travelers has rapidly increased during the last several years. 
Therefore, the economy in Japan should be considerably affected by their travel dynamics. 
For marketers in the tourism industry, for example, a global view of their travel patterns is expected to be an informative piece of information.

Human mobility data are distributed by several organizations, and an analysis of mobility patterns has been an important topic in this regard. During the past decade, in particular, datasets based on records of mobile devices \cite{Hossmann2011,Isaacman2012,Ganti2013} and geo-located tags in social networks \cite{Hawelka2013,Zagheni2014,Spyratos2019} have been actively studied. 
The target scale ranges from an intramural or urban scale to an international scale (e.g., migration).
In this paper, we describe a network analysis of travel pathways by foreign travelers in Japan. 

\subsection{Dataset}
The dataset analyzed is a set of travel records of foreign travelers in Japan that is distributed by the Ministry of Land, Infrastructure, and Transport,
called the \textit{FF-data} (Flow of Foreigners-Data) \cite{FFdata}.
In this dataset, the traveler pathways, i.e., the locations the travelers visited in chronological order, are recorded based on survey responses of the travelers themselves. 
FF-data are separated into yearly datasets; currently, datasets from 2014 through 2017 are available. 
Although these datasets contain various attributes of foreign visitors, we focus on identifying the macroscopic features of the travel pathways. 
As an interesting characteristic of such travel pathways, because Japan is surrounded by water and because the FF-data contain only the records of foreign travelers, all pathways start and end at airports or seaports.

\subsection{Travel data as a network and as a memory network}\label{MemoryNetworkMain}
It is common to characterize travel data as a network \cite{Barbosa2018}. 
A naive way to characterize travel data as a network is to represent the locations of a departure and destination as nodes and a transition of a traveler between two nodes as a directed edge. 
The result is typically a multigraph (a graph with multiple edges between a pair of nodes) because numerous travelers travel between the same pairs of locations. 

However, such a network representation has a crucial limitation as it can only encode pairwise information among the nodes. 
Although some travelers visit more than one destination during their visit, the pathways are broken down into independent transitions between each pair of locations and are encoded as edges. 
For example, although the pathway from Tokyo to Osaka, followed by Osaka to Kyoto, is quite popular, presumably for tourists, the transition from Osaka to Kyoto is recorded independent of the transition from Tokyo to Osaka. 

\begin{figure}[t!]
 \begin{center}
   \includegraphics[width= \columnwidth, bb = 0 0 253 253]{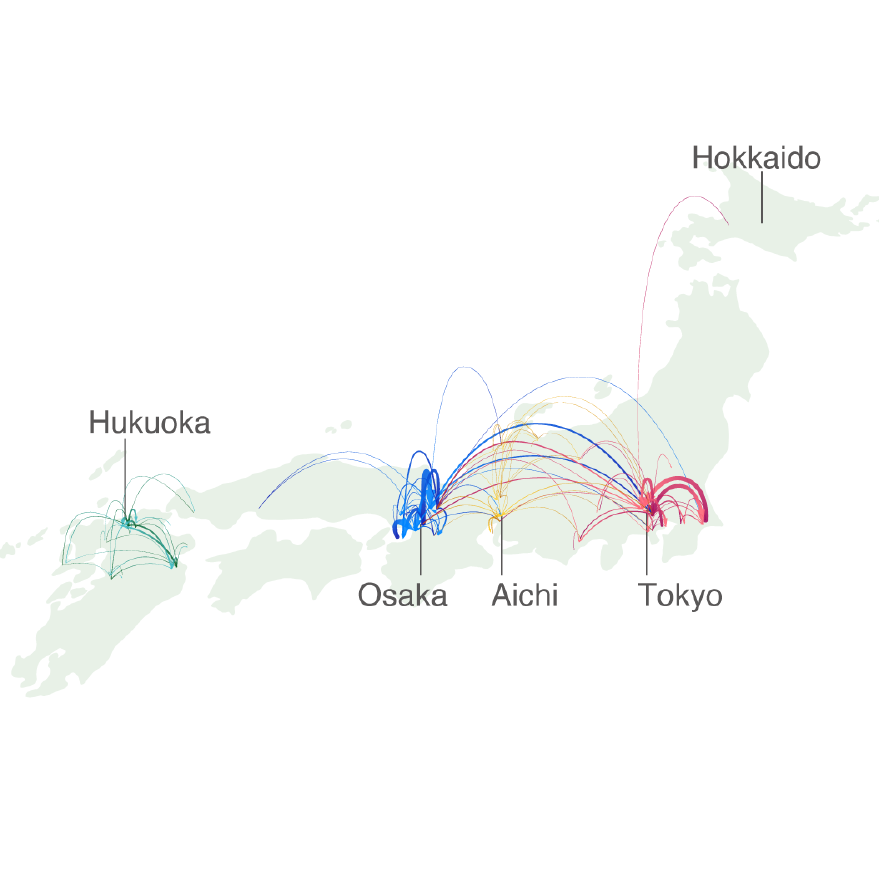}
 \end{center}
 \caption{
 Representative transitions of foreign travelers in identified modules. 
 The details of this figure are described in Sec.~\ref{sec:SecondOrderMarkov}. 
	}
 \label{fig:ModuleFeatures}
\end{figure}

We instead conducted an analysis using a higher-order network representation called a \textit{memory network} \cite{Rosvall2014,edler2017mapping,lambiotte2019networks} that incorporates the information of multiple transitions of each traveler. 
We identify densely connected components (sets of nodes) in a memory network and denote them as \textit{modules}. A structure that consists of densely connected components is termed an assortative structure, and a method to identify an assortative structure is termed community detection\footnote{In this paper, a module is loosely defined in a narrow sense. As we briefly mention in Appendix \ref{sec:MapEquation}, the community detection algorithm that we use for memory networks is not directly designed to identify densely connected components. Thus, an outcome of community detection may indicate a more general module structure. Nevertheless, we define modules as densely connected components (as a structure that we focus on). Moreover, strictly speaking, we would need to specify what we mean by ``densely connected'' exactly. When the problem is treated as an inference problem (instead of an optimization problem), we would also need to assess whether the identified modules are statistically significant. However, we treat a module as an object that is not strictly defined. As we will also mention in Sec.~\ref{sec:Stationarity}, we only use community detection to \textit{propose} a set of components that may be interpreted as modules in the sense of densely connected components. Therefore, although we refer to the identified components as modules for simplicity, it is more precise to describe them as candidates of modules. In other words, we do not define modules as an outcome of an algorithm.}. 
Figure ~\ref{fig:ModuleFeatures} shows a brief overview of the classified elements of traveler pathways. 
Note that the pathways are basically classified into modules of different regions. 
For example, however, because Tokyo is a hub for many different destinations, transitions containing Tokyo belong to many different modules. 
This is therefore an overlapping module structure. 

Herein, we provide a brief introduction to a memory network; a formal definition of the memory network can be found in ~\cite{Rosvall2014,edler2017mapping} and in Appendix \ref{MemeryNetworkFormulation}. 
An $m$th-order Markov (memory) network consists of \textit{state nodes} and the directed edges between them. 
A state node represents the destination as well as its recent $m-1$ transitions. 
For example, in a second-order Markov network, when a traveler moves from Osaka to Kyoto, we have a state node that places Kyoto as the destination with the memory of Osaka as the previous location. 
If the traveler left Tokyo before arriving at Osaka, we also have a state node that has Osaka as the destination with the memory of Tokyo, and these two state nodes are connected by a directed edge. 
Because many travelers travel along similar pathways, a memory network is typically a multigraph. 
From the viewpoint of a memory network, a simple network representation is categorized as a first-order Markov network or a memoryless network. 
The actual locations are referred to as \textit{physical nodes} in a memory network. 
The destination of a state node is regarded as a physical node to which a state node belongs. 
As a community detection algorithm for memory and memoryless networks, we employ the Infomap \cite{MapEquationURL} that optimizes an objective function called the map equation \cite{Rosvall2008,Rosvall2011} (see Appendix~\ref{MemeryNetworkFormulation} for a brief introduction of the map equation). 

\begin{figure*}[t!]
 \begin{center}
   \includegraphics[width= \columnwidth, bb = 0 0 691 329]{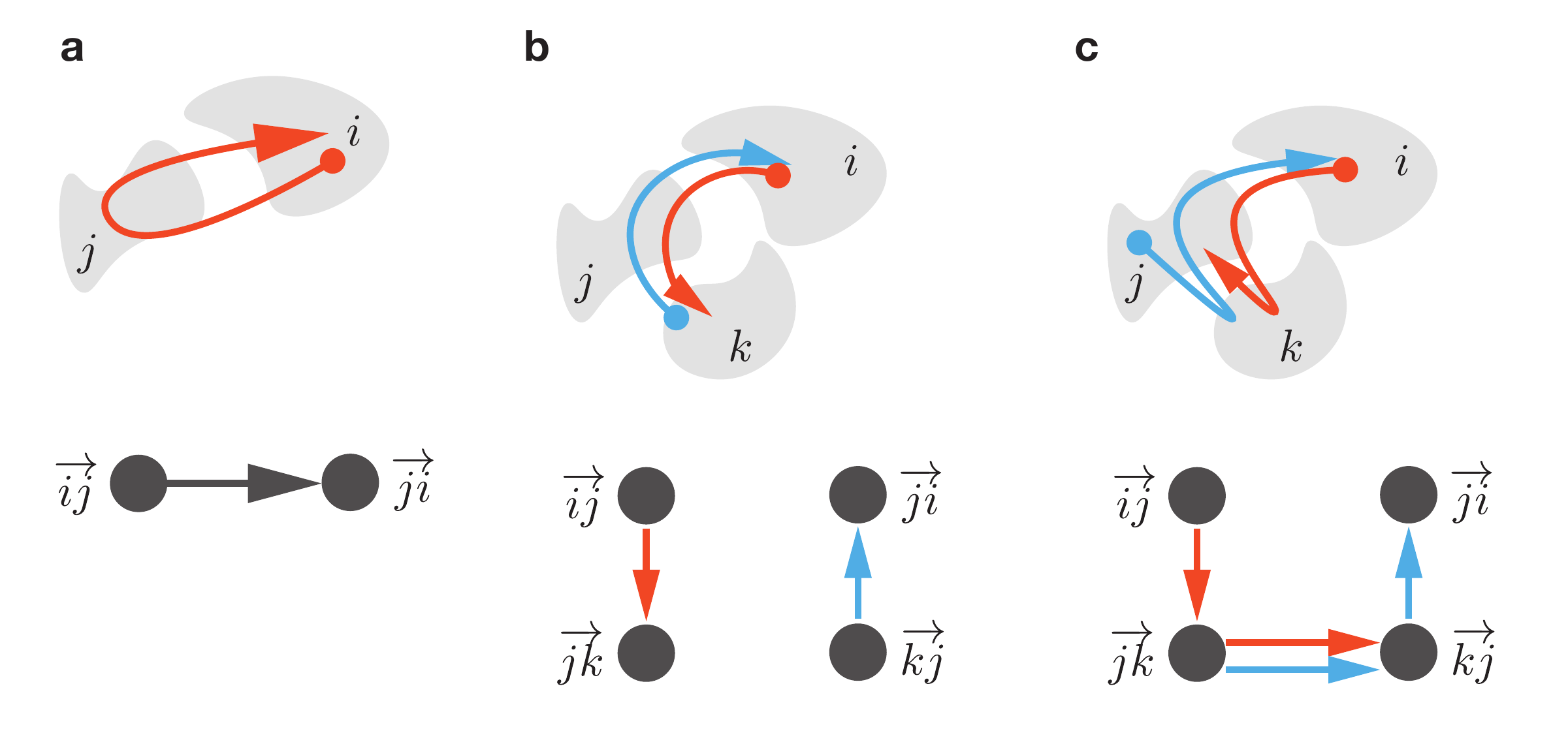}
 \end{center}
 \caption{Pathway examples (top) and the corresponding network elements in a second-order Markov network (bottom). 
	}
 \label{fig:pathway_and_memorynetwork}
\end{figure*}

To better understand the community detection on memory networks, let us consider some specific pathways in second-order Markov networks. 
We denote a state node corresponding to a transition from location $i$ to location $j$ as $\overrightarrow{ij}$. 
If a traveler moves from $i$ to $j$ and he/she moves back to $i$ again (i.e., a backtracking pathway) as shown in Fig.~\ref{fig:pathway_and_memorynetwork}{\bf a} (top), we have a directed edge from $\overrightarrow{ij}$ to $\overrightarrow{ji}$ (Fig.~\ref{fig:pathway_and_memorynetwork}{\bf a} (bottom)). 
In contrast, when we have asymmetric pathways, as shown in Fig.~\ref{fig:pathway_and_memorynetwork}{\bf b} (top), the state nodes of opposite directions are not connected (Fig.~\ref{fig:pathway_and_memorynetwork}{\bf b} (bottom)). 
Note, however, that even when the backtracking pathway is absent and pathways are asymmetric, $\overrightarrow{ij}$ and $\overrightarrow{ji}$ can be (indirectly) connected; the pathways and the corresponding network element shown in Fig.~\ref{fig:pathway_and_memorynetwork}{\bf c} is such an example. 
When the pathways in Figs.~\ref{fig:pathway_and_memorynetwork}{\bf a} and {\bf c} are popular, $\overrightarrow{ij}$ and $\overrightarrow{ji}$ are likely to belong to the same module.

It should be noted that, although representing a dataset using a memoryless network can incur a significant loss of information, it is occasionally sufficient, depending on the dataset and purpose of the analysis. 
Furthermore, even when a record of multiple transitions is not explicitly encoded in a network, it may be possible to infer similarities among nodes in some sense from such a simple network. For example, when every pair of nodes is connected by a considerably large number of edges within a certain node set, we can trivially conclude that those nodes are similar to each other. 
Therefore, before moving on to an analysis of a second-order Markov network, we first investigate the basic statistics and a first-order Markov network of the FF-data to determine whether a higher-order network representation is necessary for an extraction of a module structure.

\section{Network analysis of travel pathways}\label{sec:NetworkAnalysis}

\subsection{Basic statistics}
\begin{figure}[h!]
\begin{center}
\includegraphics[width=0.7\columnwidth, bb = 0 0 720 360]{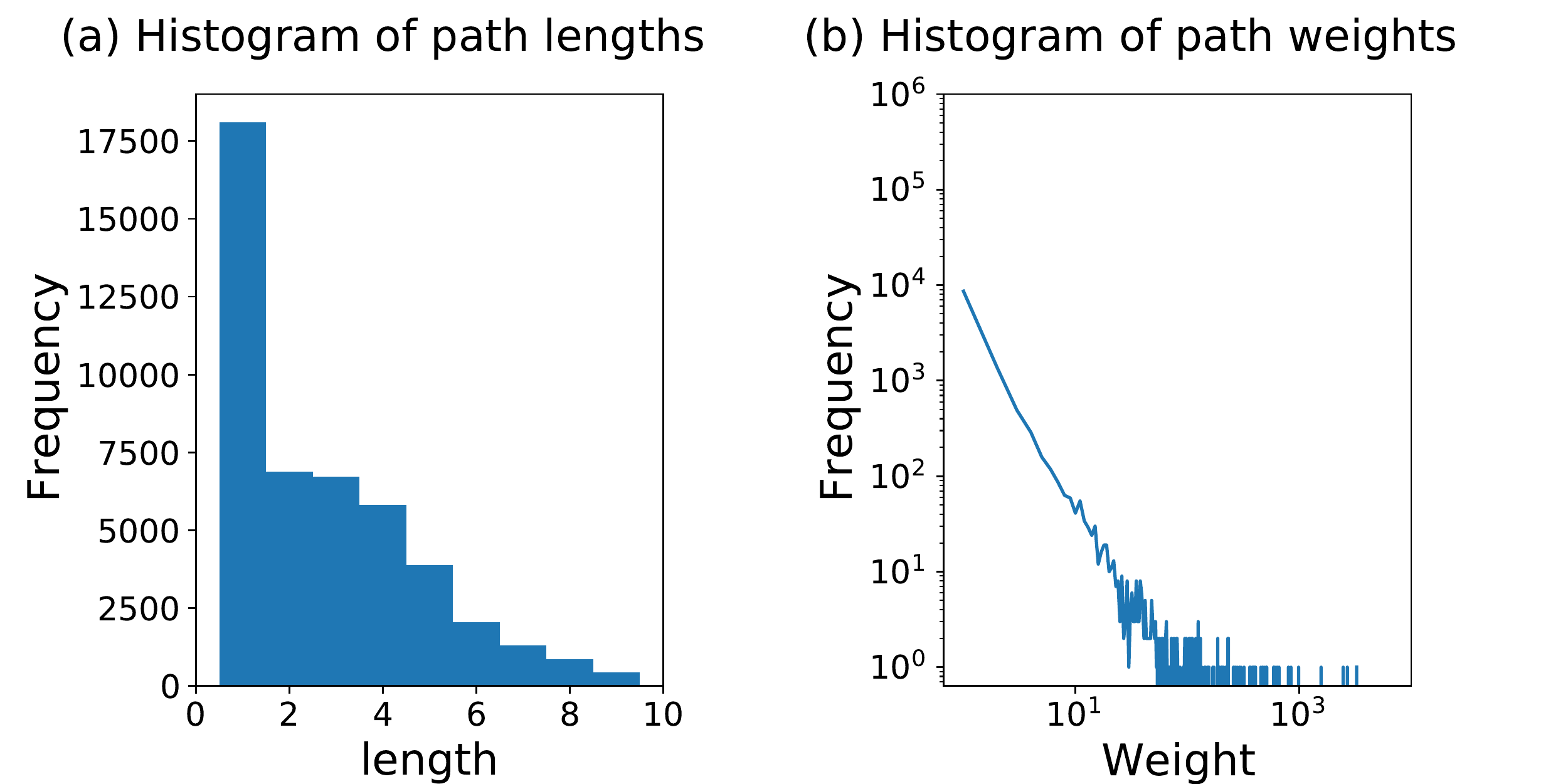}
\end{center}
\caption{Histograms of {\bf (a)} the pathway lengths (normal scale) and {\bf (b)} pathway weights (log-log scale) of the FF-data in 2017. 
In panel (b), the frequencies of the pathway weights are counted based on logarithmic binning.}
\label{fig:distribution}
\end{figure}

Let us first look at the basic statistics of the FF-data. 
In Table \ref{tab:basicstatistics}, we summarized the numbers of destinations (including the airport and seaport first visited), travelers, unique pathways (i.e., itineraries), and unique edges (first and second order, respectively), as well as the median, lower and upper quartiles of the pathway lengths (number of steps of transitions), and weights. 
The distributions of the pathway lengths and weights are shown in Fig.~\ref{fig:distribution}. 
We define the weight of a pathway as the number of travelers who traveled along exactly the same pathway. (This should not be confused with the weight of a memory network, which is defined as the number of transitions occurring between two state nodes.) 
It is confirmed that both distributions are heavily tailed and a considerable number of travelers travel to more than two destinations. 
This observation implies that there is no evident cutoff length and weight of the pathways or a characteristic scale of the length and weight that we should consider. 
In other words, apparently, multiple pathway lengths significantly contribute on pathway structures.
Therefore, the first-order Markov network may be inadequate for a correct understanding of travel pathways because it only takes into account the one-step transitions. 

\begin{table*}[t!]
\begin{center}
\caption{Basic statistics of travel pathway in the FF-data for each year. 
Each triple in the pathway lengths and weights represent the 25\% quantile, median, and 75\% quantile, respectively.}
\begin{tabular}{@{}lcccc}
\toprule
Year                      & 2014 & 2015 & 2016 & 2017 \\ \midrule
Num. of destinations         &    80  &      82&      83&      84\\
Num. of travelers              &   46,635   &     60,640&     63,116 &     65,191\\
Num. of pathways  &10,887 & 12,093& 11,995 & 11,818 \\
Num. of edges (1st order)  &1,995 & 2,130& 2,121 & 2,105 \\
Num. of edges (2nd order) & 9,676 &10,679 & 10,937 & 10,760 \\
Pathway lengths       &   (2.0, 3.0, 5.0)  &     (2.0, 3.0, 5.0)&     (2.0, 3.0, 5.0)&      (3.0, 4.0, 6.0) \\
Pathway weights &   (1.0, 1.0, 1.0)   &      (1.0, 1.0, 2.0) &     (1.0, 1.0, 2.0) &      (1.0, 1.0, 2.0)  \\ \bottomrule
\end{tabular}
\label{tab:basicstatistics}
 \end{center}
\end{table*}

\subsection{First-order Markov network}\label{sec:FirstOrderMarkov}

\begin{figure}[t!]
\begin{center}
\includegraphics[bb = 0 0 827 767,  width=\columnwidth]{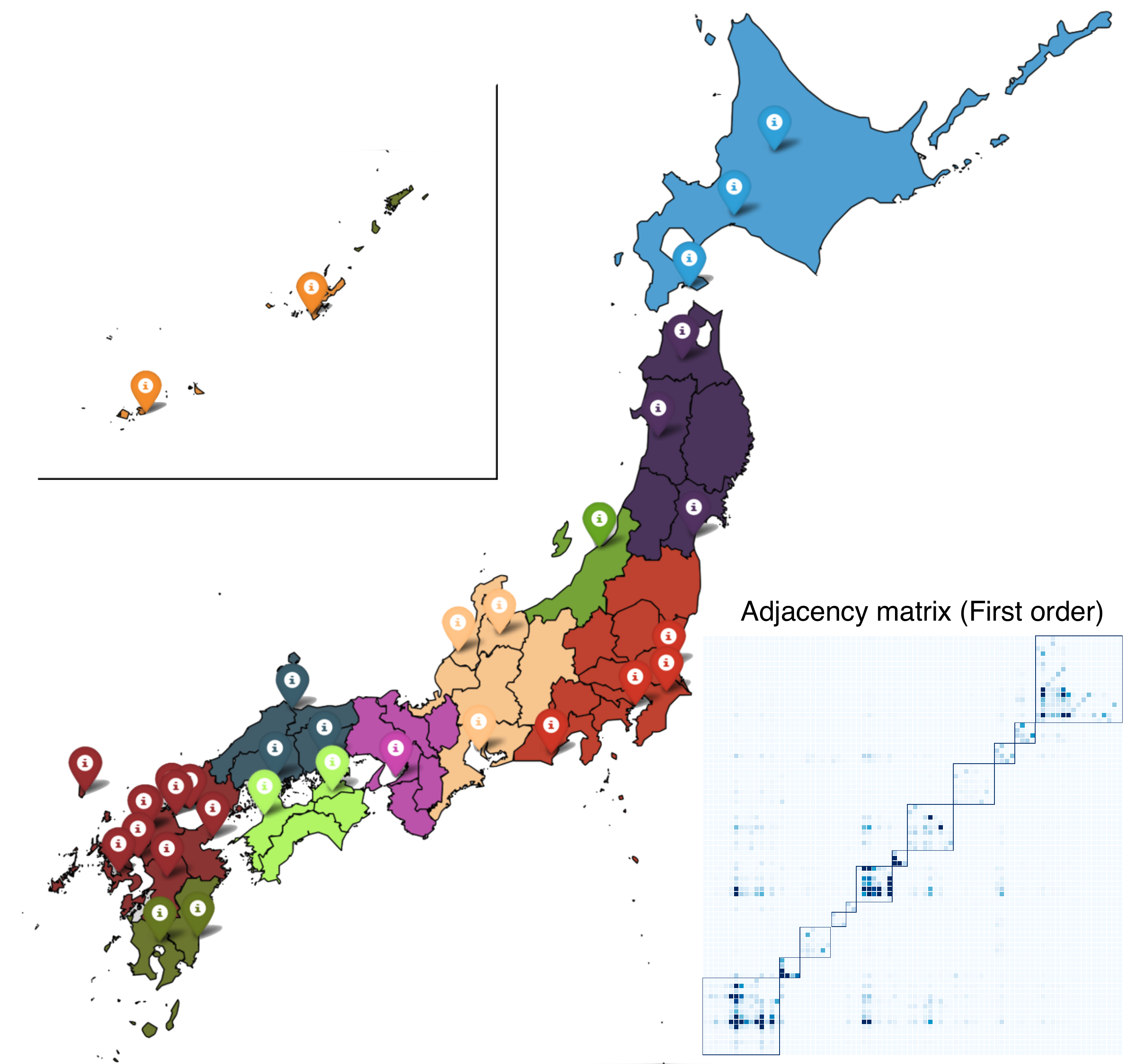}
\end{center}
\caption{Results of community detection on a memoryless network using the Infomap: {\bf (Main part)} Classification on the map figure (the upper-left part represents Okinawa islands) and {\bf (Lower-right part)} adjacency matrix of the network sorted according to the module indices. In the map figure, the pin icons indicate the locations of airports and seaports in the dataset.} 
\label{fig:partition_1st}
\end{figure}

We then consider the community detection of the first-order Markov network. 
Here, we focus on the dataset of 2017. 
Figure \ref{fig:partition_1st} shows the partitioning of 47 prefectures, 33 airports, and 4 seaports in Japan. 
The corresponding adjacency matrix is shown at the bottom right; each element is a node pair, and the color depth indicates the weight of an edge. 
Note that the FF-data includes transitions within a prefecture, indicating self-loops in the first-order Markov network. 
The order of the node labels is sorted based on the modules identified using the Infomap and each module is specified by a blue box. 
It is confirmed that the Infomap identifies an assortative module structure on the first-order Markov network. 
The modules apparently correspond to the geographical regions in Japan: Hokkaido, Tohoku, Chyuetsu, Kanto, Chubu, Kansai, Chugoku, Shikoku, Northern Kyushu, and Southern Kyusyu with Okinawa regions. 

Although the present results may sound reasonable, there is an important concern. 
Notice that all airports and seaports belong to the module of the region where the airport or seaport is located. 
This observation implies that the present result is simply dominated by the transition from--to airports and seaports. 
If this is the case, then the result may not reflect any informative travel patterns. 
This is further confirmed because the result of community detection is considerably varied by removing the airport and seaport nodes. 
Moreover, we have also confirmed that the module boundaries of the first-order Markov network can be reproduced with high accuracy by a simple label aggregation based on the module labels of the airport nodes. 
These results are shown in Appendix \ref{FirstOrderAnalysisDetail}. 
Based on this, we move to an analysis of the second-order Markov networks.

\subsection{Second-order Markov network}\label{sec:SecondOrderMarkov}

\begin{figure}[t!]
\begin{center}
\includegraphics[width=\columnwidth, bb = 0 0 1273 1166]{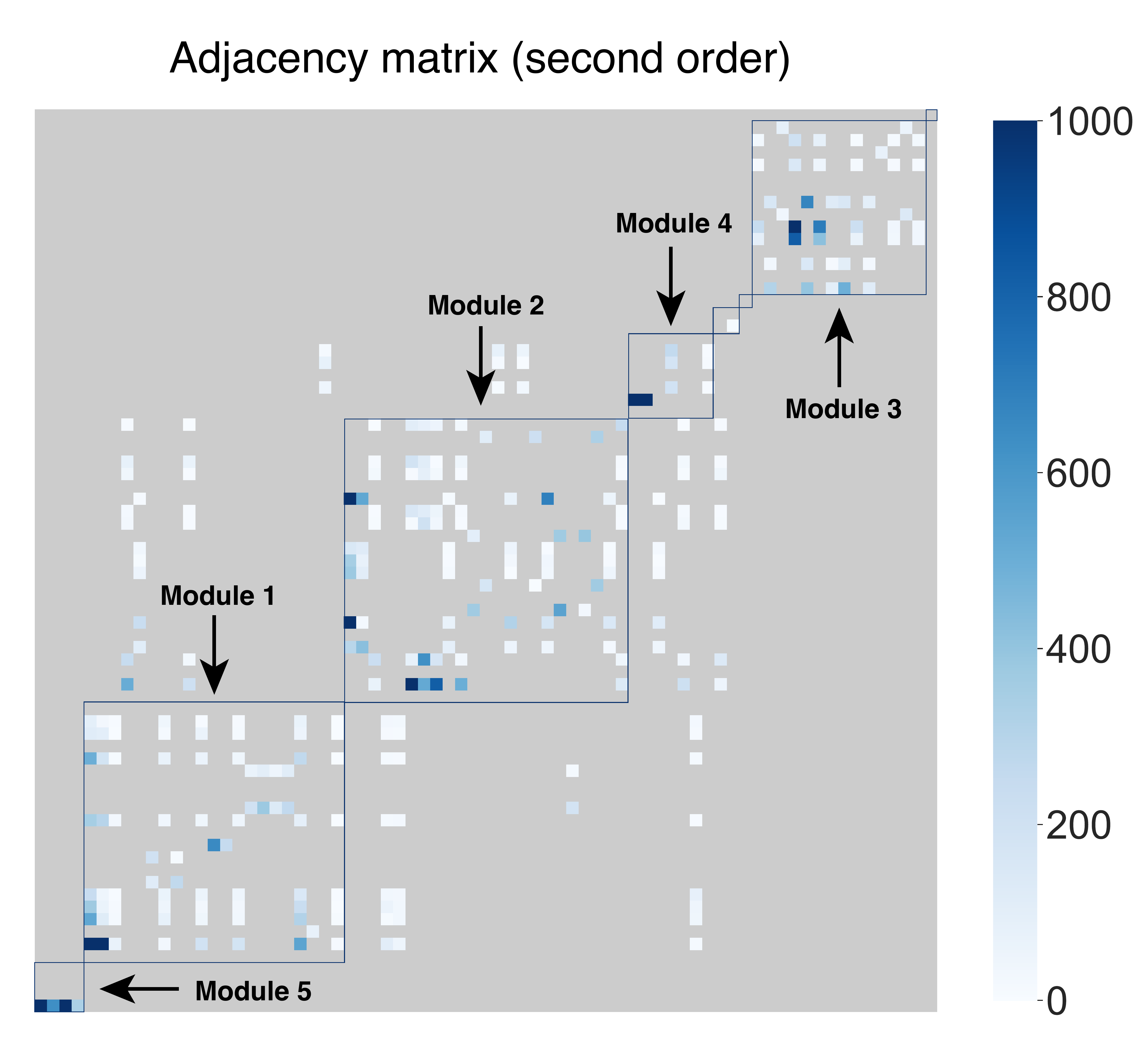}
\end{center}
\caption{Adjacency matrix of the second-order Markov network. 
Each element represents a state node, and the color depth indicates the weight of an edge in the network. 
The gray elements represent the forbidden edges. 
The order of node labels is sorted based on the module assignments, and the identified modules are specified by blue boxes.}
\label{fig:SecondOrderAdjacencyMatrix}
\end{figure}

Here, we classify the state nodes of the second-order Markov networks into modules using the Infomap, which can be directly applied to higher-order Markov networks. 
Figure \ref{fig:SecondOrderAdjacencyMatrix} shows the adjacency matrix of a subgraph in the second-order Markov network corresponding to the dataset in 2017. 
The state nodes in the subgraph are extracted such that each state node belongs to one of the eight largest modules and has a total edge weight to other state nodes (i.e., degree) larger than 300.
The matrix elements are sorted based on the module assignments. 

Note that there are forbidden edges in the second-order Markov networks. 
For example, when an edge is present from state node $\alpha$ to state node $\beta$, an edge in the opposite direction is prohibited because such an edge does not constitute a pathway, e.g., for $\alpha = \mathrm{Tokyo}\rightarrow\mathrm{Osaka}$ and $\beta = \mathrm{Osaka}\rightarrow\mathrm{Kyoto}$, $\beta$ to $\alpha$ is not possible because a traveler cannot be at Kyoto and Tokyo at the same time. 
Analogously, the edges of both directions are prohibited when neither of them constitutes a pathway. 
The elements corresponding to the forbidden edges are indicated in gray in Fig.~\ref{fig:SecondOrderAdjacencyMatrix}. 

It is confirmed that the Infomap identifies an assortative structure on top of the constraint of the forbidden edges. 
Larger-sized modules exhibit assortative structures, even though the edges to different modules are not strictly prohibited. 
In contrast, the modules of a relatively smaller size are identified owing to a severe constraint of the forbidden edges. 

Let us then explore the identified module structure in more detail. 
Figure \ref{fig:ModuleFeatures} shows popular transitions that appear in the four largest modules. 
The arrows of the same color represent transitions within the same module. 
The color depth of an arrow distinguishes the prior (lighter color) and destination (deeper color) physical nodes. 
The thickness of each arrow reflects the weight of a transition measured based on the \texttt{flow} (PageRank) in the outcome of the Infomap. 

\begin{figure*}[t!]
 \begin{center}
   \includegraphics[width= \columnwidth, bb = 0 0 740 529]{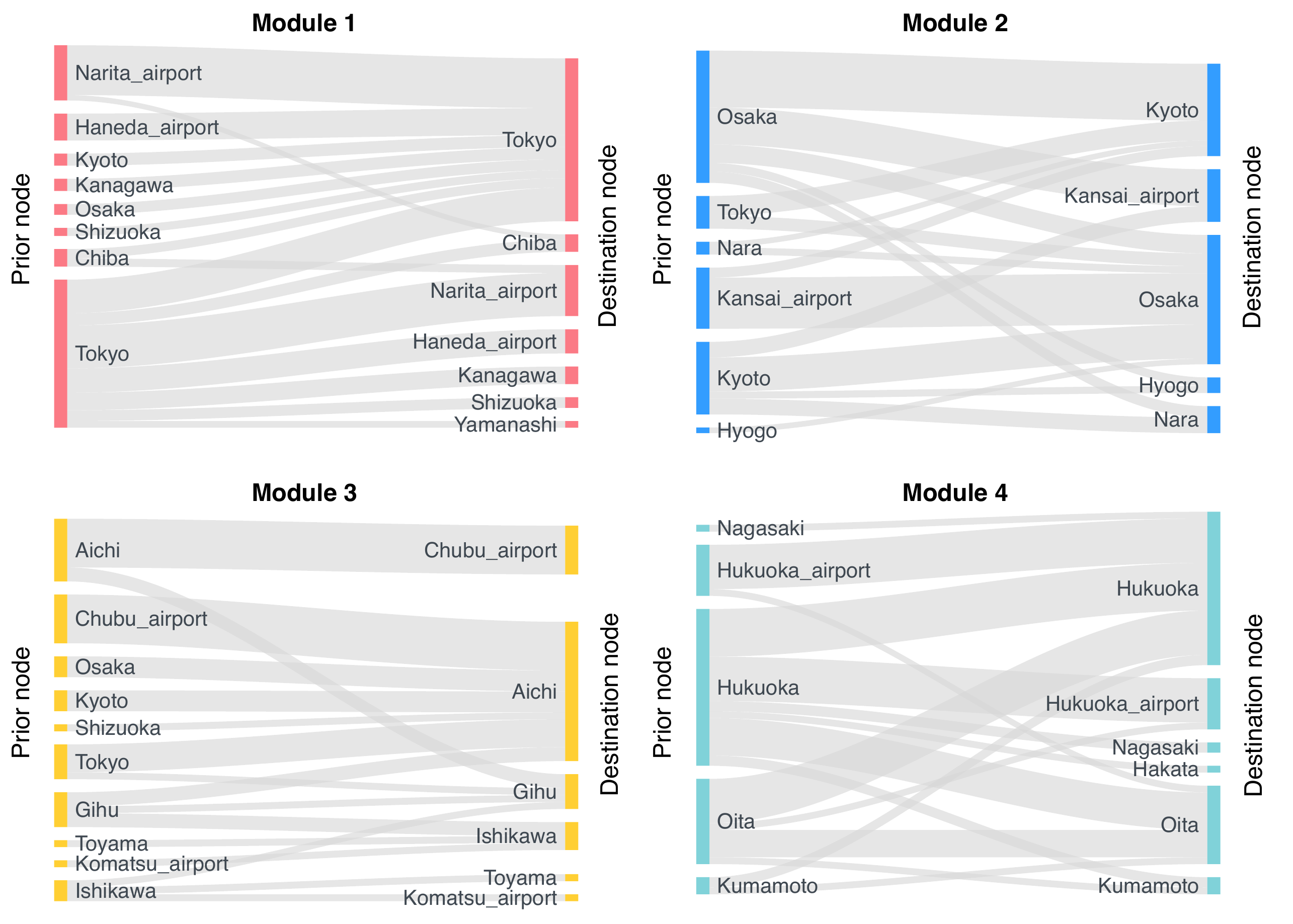}
 \end{center}
 \caption{
 Sankey diagrams of the four largest modules in the second-order Markov network. 
 The size of a strip of each location corresponds to the relative amount of flow from--to the location in the memory network. 
	}
 \label{fig:SankeyDiagrams}
\end{figure*}

We show the Sankey diagrams of the four largest modules in Fig.~\ref{fig:SankeyDiagrams}. 
The locations on the left-hand side of each diagram represent the physical nodes visited before arriving at the (physical) destination nodes; we refer to them as prior nodes. 
Conversely, the locations on the right-hand side of each diagram represent the destination nodes. 
It is confirmed that, for example, the transition from Kyoto to Tokyo appears in Module 1, while the transition from Tokyo to Kyoto appears in Module 2 (see Fig.~\ref{fig:pathway_and_memorynetwork} and the corresponding part in Sec.~\ref{MemoryNetworkMain} for the implication).
It is also confirmed that the prior nodes and the destination nodes are strongly asymmetric in Module 3. 
In contrast, the prior nodes and destination nodes in Modules 2 and 4 are relatively symmetric. 
These observations imply that the visits of locations in Module 3 are direction-sensitive, while those in Modules 2 and 4 are less direction-sensitive.

\section{Stationarity of the modules}\label{sec:Stationarity}
The readers may wonder if the identified modules in the second-order Markov network are ``significant.''
It should be noted here that in this study, we did not perform community detection as a task of statistical inference. 
That is, we only used the Infomap as a tool to propose several informative node sets that exhibit an assortative structure.

\begin{figure}[t!]
\begin{center}
\includegraphics[width=\columnwidth, bb =0 0 842 595]{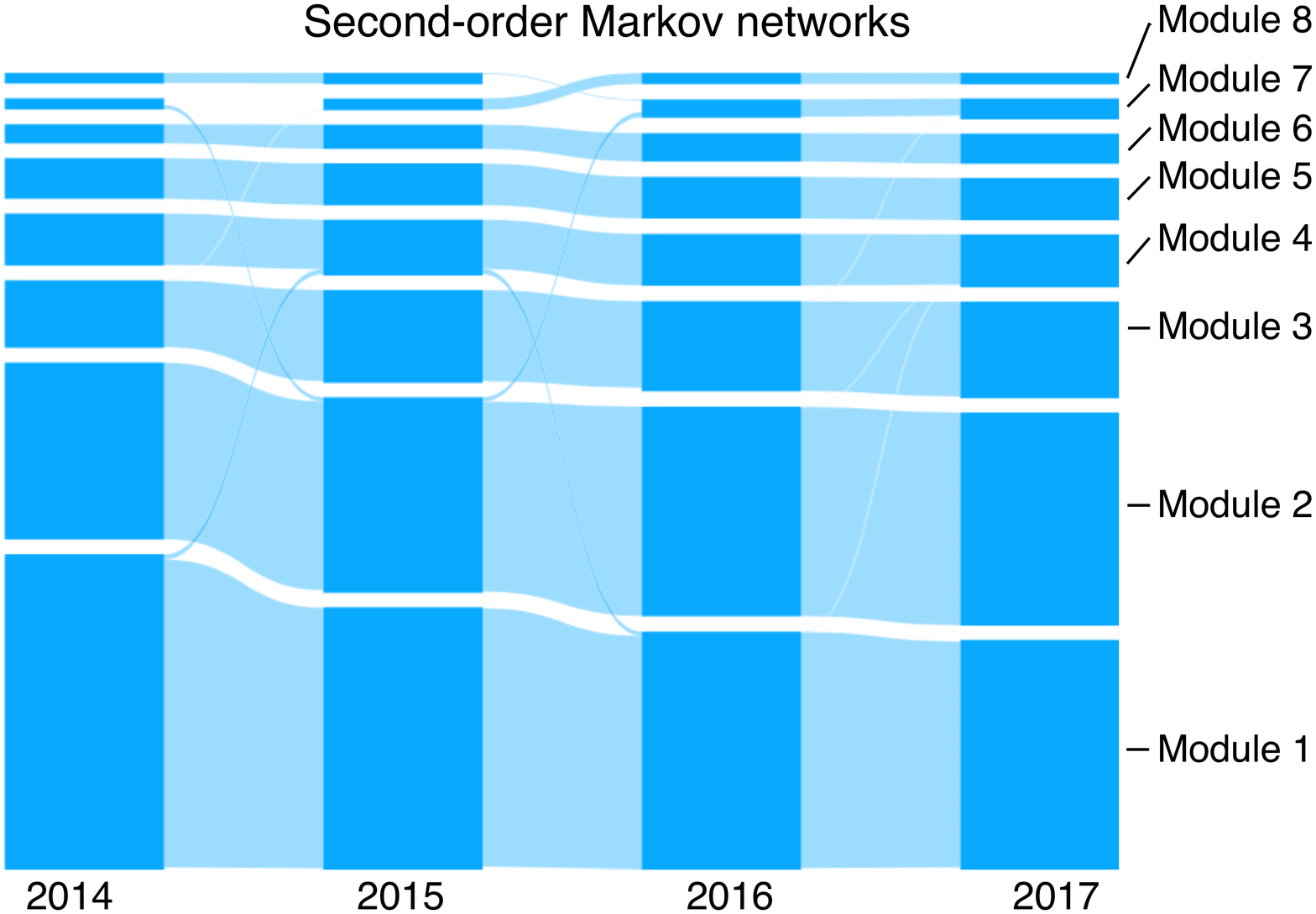}
\end{center}
\caption{Sankey diagram comparing the modules identified in the second-order Markov networks of successive years.}
\label{fig:alluvial_diagram2}
\end{figure}

However, we still need to assess which of the identified modules are informative. 
For example, a number of small modules are identified as a result of the Infomap (although we only show large modules in this paper), and it is doubtful whether they are worth further consideration. 
These small modules might have appeared because of the algorithmic infeasibility of the optimization algorithm, or they might have emerged because of the random nature of the network. 
In fact, by comparing the randomness assumed in the map equation \cite{smiljanic2019mapping}, a statistical assessment of significance should be possible. 
However, we should also note that, because the memory network of the FF-data must be strongly affected by the geographic constraint and the use of airports, the plausibility of the statistical model is also a considerable issue. 

Therefore, instead of assessing the statistical significance, we assess the significance of the identified modules based on the consistency \cite{Kawamoto2018}. 
By using the Sankey diagram\footnote{We used Alluvial Generator \cite{MapEquationURL} to draw Fig.~\ref{fig:alluvial_diagram2}. Alluvial Generator generates an alluvial diagram \cite{Rosvall2010}, which is equivalent to the Sankey diagram when there are no degrees of freedom of color depth.}, we assess whether the modules are stationary in the FF-data across different years. 
Because it is unlikely that the module structures largely differ in reality, if it happens, we should conclude that the modules are simply identified out of noises. 
Figure \ref{fig:alluvial_diagram2} indicates that, for larger modules among the ones focused upon in Sec.~\ref{sec:SecondOrderMarkov}, nearly the same modules are indeed identified in all datasets. 
From these observations, we conclude that the large modules considered in Sec.~\ref{sec:SecondOrderMarkov} are informative\footnote{Of course, there is a possibility that modules are stationary even when the algorithm only captures noise. This assessment is still very loose. However, the result here indicates that we cannot positively conclude that the identified modules are only due to noise.}.

\section{Conclusion}
Based on the observation that the FF-data apparently have structures that cannot be explained by a simple network representation, we revealed the module structure of the travel pathways of foreign visitors using the framework of the memory network and the map equation (and the Infomap as the specific algorithm). 

The results herein offer a better understanding of the travel patterns of foreign visitors. When travel patterns are utilized for deciding government policies or marketing research in the tourism industry, it is important to recognize whether a travel to/from a particular location is considerably conditioned on the location visited previously/later. It would be difficult to identify such a feature by skimming through the raw FF-data. Moreover, for the pathways that are highly correlated, the results of memoryless network approaches are not very reliable. The present analysis based on the second-order Markov networks successfully revealed dependencies on the previously/later visited locations. 

We expect that the results described herein will be even more useful by relating the identified modules with regional factors. For example, it is interesting to investigate how the modules are correlated to the amount of consumption by tourists and the degree of development of public transportation in each prefecture. 

It is difficult to forecast whether the structures we found will exist in the future, particularly during an era in which the way people live and travel is drastically changing. 
In any case, however, we expect that our results will serve as a reference for future data analysis of traveling patterns.

\begin{acknowledgements}
This work was supported by the New Energy and Industrial Technology Development Organization (NEDO). 
\end{acknowledgements}


\appendix

\section{Memory networks and the map equation}\label{MemeryNetworkFormulation}
In this section, we briefly review the formulations of the memory network \cite{Rosvall2014,edler2017mapping} and the map equation \cite{Rosvall2008,Rosvall2011} used to analyze the FF-data.

\subsection{Memory networks}\label{sec:MemoryNetwork}
A memory network is a higher-order network representation that incorporates information of multi-step transitions, or pathways. 
Given a set of pathways as inputs, the memory networks are constructed as a directed \footnote{Although the direction is already encoded between a pair of states, edges also have directions in the memory network.} weighted network. 
Throughout this article, we assume that the pathways are formed as the result of non-Markovian (i.e., stochastic) dynamics, even if the actual pathways are not strictly constructed in a stochastic manner.

\subsubsection{Definitions and notations}
\begin{figure*}[t!]
 \begin{center}
   \includegraphics[width= \columnwidth, bb = 0 0 880 289]{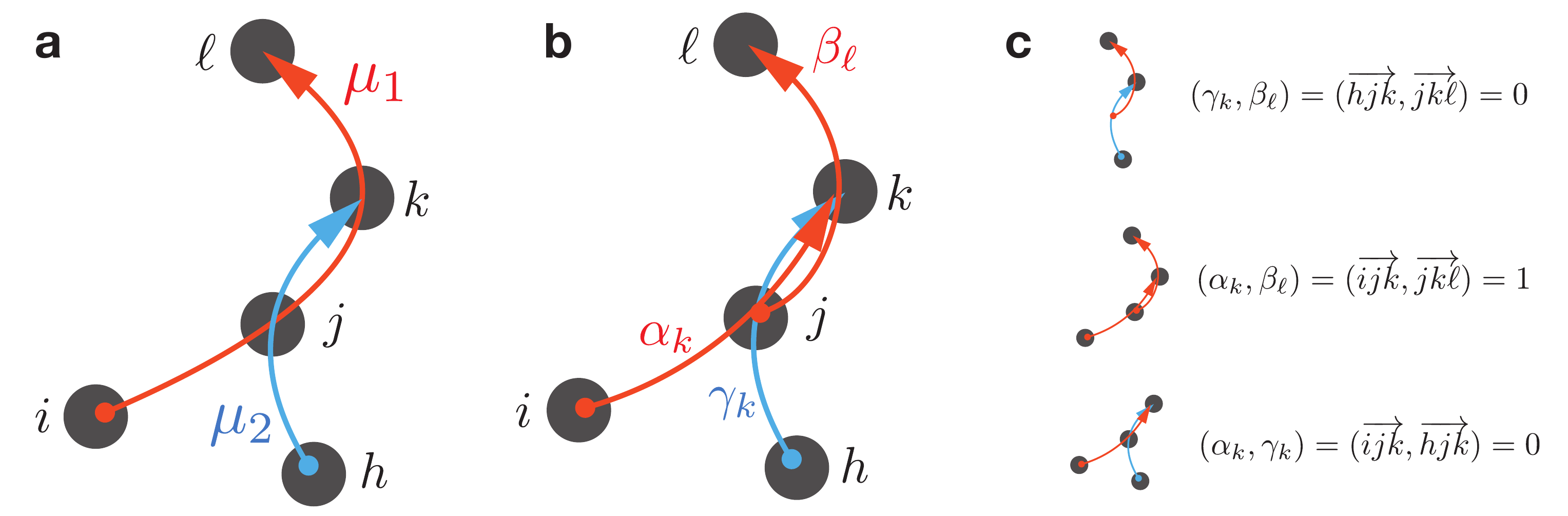}
 \end{center}
 \caption{
 Example of memory network of third-order Markov dynamics of two pathways on four physical nodes. 
  The pathways ({\bf a}) $\mu_{1}$ and $\mu_{2}$ are decomposed into ({\bf b}) $\alpha_{k}$, $\beta_{\ell}$, and $\gamma_{k}$. 
  The adjacencies of pairs of state nodes are shown in ({\bf c}). 
	}
 \label{fig:MemoryNetworkSchematic}
\end{figure*}

As mentioned in Sec.~\ref{MemoryNetworkMain}, the actual nodes (locations) are referred to as \textit{physical nodes} in contrast to the \textit{state nodes}, which are considered in the memory network. 
Following the convention in Refs.~\cite{Rosvall2014,edler2017mapping}, we used Roman characters for physical nodes and Greek characters for state nodes.
We denote a pathway, i.e., a history of visited physical nodes, as follows. 
\begin{align}
\mu = \overrightarrow{ijk\ell}. 
\end{align}
This indicates that a pathway labeled $\mu$ is a record in which transitions have occurred in order of $i \to j \to k \to \ell$. 
A physical node in a pathway is called a \textit{state}. 
Example pathways on a few physical nodes are shown in Fig.~\ref{fig:MemoryNetworkSchematic}. 

{\bf \textit{State nodes.}} 
We call a subset of a pathway a \textit{path}. 
The state node is a node that represents a path of a specified length. 
For example, 
\begin{align}
\alpha_{k} = \overrightarrow{ijk} \label{PathExample}
\end{align}
is the state node corresponding to the path that has $k$ as the physical destination node in the pathway $\mu$. 
The length of a path recorded in a state node depends on the order of non-Markovianity considered. 
In a memory network of $m$th-order Markov dynamics, a set of state nodes consists of paths with $m-1$ step transitions. 

Note that a state node is a unique object with respect to the series of states; that is, if multiple pathways have a common path $\overrightarrow{ijk}$, it is uniquely denoted as $\alpha_{k}$ irrespective of the pathway to which the path belongs. 
Note also that the path $\alpha_{k}$ is indexed by $\alpha$, and not $k$. The role of subscript $k$ is to clarify the correspondence between the state node and the physical destination node; as the interpretation of $\overrightarrow{ijk}$, \textit{the state node exists in physical node $k$ with the memory of the previously visited nodes $i$ and $j$.}

{\bf \textit{Edges.}} 
A pair of state nodes are connected by a directed edge in a memory network when the last $m-1$ states of one state node are equal to the first $m-1$ states of the other state node in a single pathway; that is, a set of subsequent paths is connected. 
For example, in third-order Markov dynamics, shown in Fig.~\ref{fig:MemoryNetworkSchematic}, state nodes $\beta_{k}$ and $\beta_{\ell}$ are connected by a directed edge, whereas the other paths are not. 

\subsection{The map equation}\label{sec:MapEquation}
The map equation \cite{Rosvall2008,Rosvall2011} is a popular community detection method that utilizes the minimum average description length of a random walker on a network. 
The map equation is a framework that takes the transition rates between nodes as input and aims to identify the optimum module structure. 

Roughly speaking, the map equation uses a hierarchical encoding for the description of a random walker trajectory. 
Instead of simply using the node labels to record the movement of a random walker, the map equation uses hierarchical module labels. 
By optimally selecting the hierarchical module labels on nodes and transitions, we can obtain an efficient description of the trajectory of a random walker. 
The optimum module labels are interpreted as the result of community detection in the map equation. 
Importantly, we do not need to actually allow a random walker to run on a network or specify the coding algorithm to describe the walker?s trajectory. 
In the stationary limit, for a given module labeling, the minimum description length\footnote{Note that ``minimum'' here indicates the minimum among all possible encoding algorithms and should not be confused with the minimum among all possible module labeling. The map equation aims to identify the minimum of the average minimum description length.} per unit time can be calculated analytically using the transition rates between nodes and modules. 

Although the map equation was originally formulated as a community detection method for memoryless dynamics on a network, using the memory network representation, the applicability of the map equation can be directly extended to the case of non-Markov dynamics. 

\section{Further analysis of the first-order Markov network}\label{FirstOrderAnalysisDetail}
\begin{figure*}[t!]
\begin{center}
\includegraphics[width=\columnwidth, bb = 0 0 842 595]{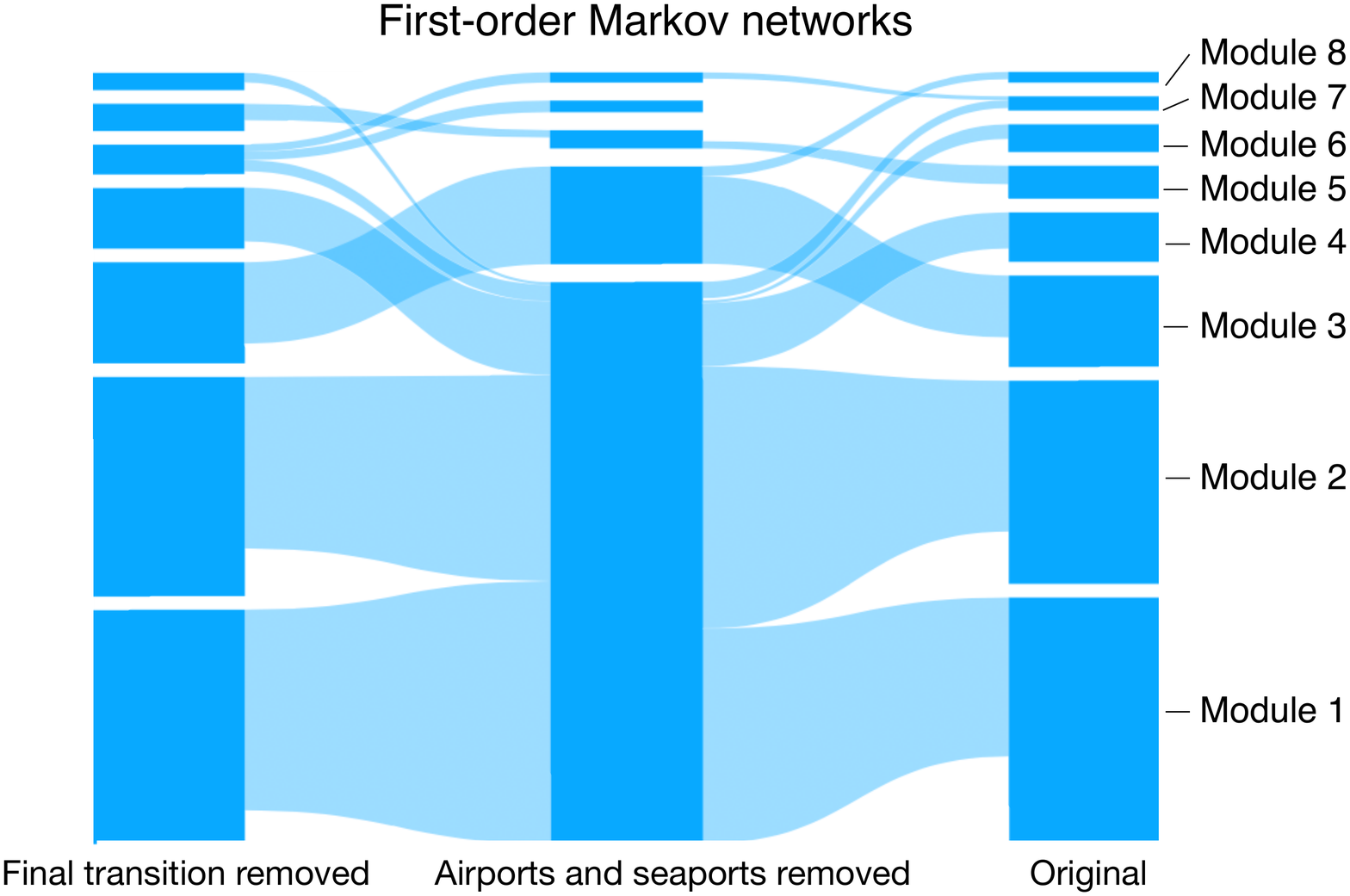}
\end{center}
\caption{Sankey diagram comparing the eight largest modules identified in the first-order Markov network of the dataset in 2017. 
The modules on the left represent the result in which we removed the final transition of each traveler.
The modules in the middle represent the result in which we removed the airports and seaports in the dataset. 
The modules on the right represent the result that we showed in the main text. 
}
\label{fig:1st_order_comparison}
\end{figure*}

In this section, we investigate the plausibility of our hypothesis that the modules identified by the Infomap on the first-order Markov network are dominated by airports and seaports. 
To this end, we compare the community detection of the following networks with the original one: 
\begin{itemize}
\item The first-order Markov network in which we removed the airports and seaports in the dataset
\item The first-order Markov network in which we removed the final transition of each traveler (because all travelers finally move toward an airport or seaport\footnote{We could remove the initial transitions instead of the final ones. In either case, the connectivity between two nodes will be affected if they are connected via airports and seaports.})
\end{itemize}
The right part of the Sankey diagram\footnote{Again, we used Alluvial Generator.} in Fig.~\ref{fig:1st_order_comparison} shows that the modules identified in the network without airport and seaport nodes are considerably distinct from those of the original first-order Markov network. 
The left part of the Sankey diagram shows that the module assignments are still distinct even if we remove the effect of transition to the airports and seaports at the end of the pathways. 
The result here is not a proof that the modules in the first-order Markov networks are not informative. 
However, it is evident that the modules are considerably affected by airports and seaports. 

To further confirm our hypothesis, we show that the module boundaries of the first-order Markov network can be reproduced with high accuracy by the following simple label aggregation. 
For each node corresponding to an airport, we assign the module label obtained by the Infomap. 
Then, for other nodes corresponding to the cities and prefectures, we assign the same module label as that of the neighboring airport node with the largest edge weight (every city/prefecture node is connected to at least one airport node). 
Although the nodes corresponding to the seaports are neglected in this label aggregation, their contribution is negligible even if they are included. 

\begin{figure*}[t!]
\begin{center}
\includegraphics[width=\columnwidth, bb = 0 0 973 870]{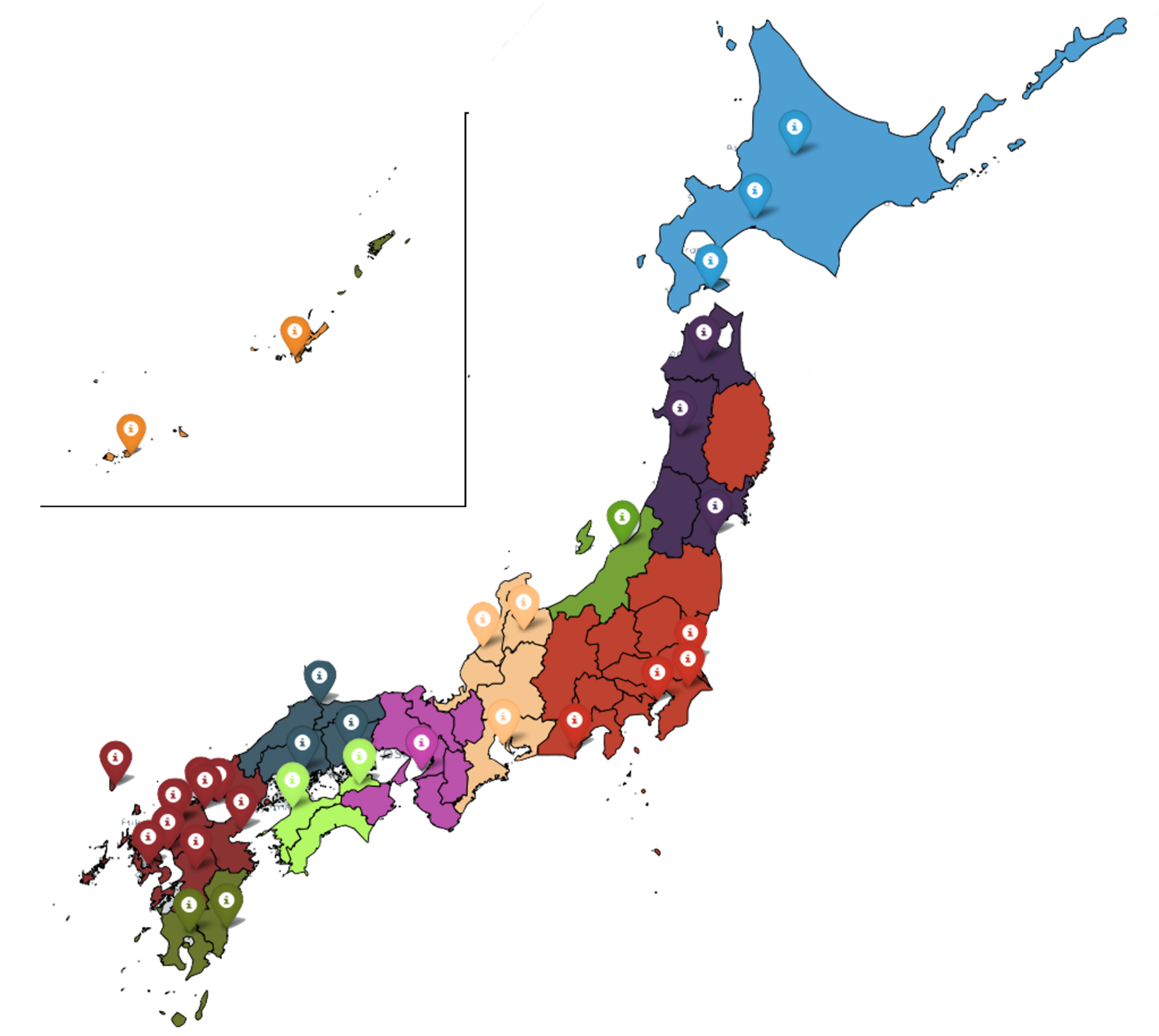}
\end{center}
\caption{
Modules obtained by a simple label aggregation. 
}
\label{fig:aggregated_around_airports}
\end{figure*}

Figure \ref{fig:aggregated_around_airports} shows the result of the aforementioned label aggregation. 
Indeed, except for three prefectures (Iwate, Nagano, and Tokushima), we obtained the same module assignments as those of the Infomap \footnote{For the pin icons corresponding to the seaports in Fig.~\ref{fig:aggregated_around_airports}, we assigned the module label obtained by the Infomap.}. 
Note that this label aggregation itself is not a community detection algorithm, because we need to specify the module assignments of the airport nodes. 
The present analysis indicates, however, that the behavior of the community detection on the first-order Markov network can be well-described by our hypothesis.

 \section*{Data availability}

The results of community detection described in this paper are available at \cite{HashimotoGithub}.

%
%

\bibliographystyle{spphys}       
\bibliography{ref_ffdata}   

%
%

\end{document}